\documentclass[aps,prl,twocolumn,showpacs,superscriptaddress]{revtex4-2}
\usepackage{latexsym}
\usepackage{amssymb}
\usepackage{graphicx}
\usepackage{amsmath}
\usepackage{bm}
\usepackage[colorlinks,
          linkcolor=black,
            citecolor=black,
            urlcolor=blue
           ]{hyperref}
\usepackage{verbatim}
\usepackage{mathrsfs}
\usepackage{extarrows}
\usepackage{comment}
\usepackage{mathtools,slashed}
\usepackage{ulem}

\usepackage{cancel}

\begin{document}

\title{Exact quantization of topological order parameter in SU($N$) spin models, $N$-ality transformation and ingappabilities}

\author{Hang Su}
\affiliation{Institute of Condensed Matter Physics, School of Physics and Astronomy, Shanghai Jiao Tong University, Shanghai 200240, China}
\author{Yuan Yao}
\email{smartyao@sjtu.edu.cn}
\affiliation{Institute of Condensed Matter Physics, School of Physics and Astronomy, Shanghai Jiao Tong University, Shanghai 200240, China}
\author{Akira Furusaki}
\affiliation{Condensed Matter Theory Laboratory, RIKEN CPR, Wako, Saitama 351-0198, Japan}
\affiliation{Quantum Matter Theory Research Group, RIKEN CEMS, Wako, Saitama 351-0198, Japan}

\begin{abstract}
We show that the ground-state expectation value of twisting operator is a topological order parameter for $\text{U}(1)$- and $\mathbb{Z}_{N}$-symmetric symmetry-protected topological (SPT) phases in one-dimensional ``spin'' systems --- it is quantized in the thermodynamic limit and can be used to identify different SPT phases and to diagnose phase transitions among them. 
We prove that this (non-local) order parameter must take values in $N$-th roots of unity, and its value can be changed by a generalized lattice translation acting as an $N$-ality transformation connecting distinct phases.
This result also implies the Lieb-Schultz-Mattis ingappability for SU($N$) spins if we further impose a general translation symmetry. 
Furthermore, our exact result for the order parameter of SPT phases can predict a large number of LSM ingappabilities by the general lattice translation.
We also apply the $N$-ality property to provide an efficient way to construct possible multi-critical phase transitions starting from a single  Hamiltonian with a unique gapped ground state.
\end{abstract}

\date{\today}


\maketitle

\paragraph{Introduction.---}
Distinguishing and recognizing various quantum phases is a central issue in condensed matter physics and statistical physics, but it is a difficult task in strongly interacting systems.
Symmetry often plays an essential role in the classification of quantum phases,
and the pattern of spontaneous symmetry breaking (SSB) determines the relationship between
different phases in the Landau-Ginzburg-Wilson (LGW) paradigm \cite{Landau:1937aa}.
Recently, significant progress beyond the LGW approach has been made in the study
of topological phases of matter, in particular, in the context of 
symmetry-protected topological (SPT) phases \cite{Gu:2009aa,Chen:2010aa,Pollmann:2012aa,Wen:TOreview2013,Duivenvoorden:2013aa},
where symmetry is not broken and local order parameters are not available.
SPT phases are stable as long as the symmetry is preserved, and a phase transition between two distinct SPT phases occurs
when the energy gap above the unique ground state is closed.
The universality class of such phase transitions reflects the topology of the quantum phase diagram,
and a generalized concept of order parameters for SPT phases will also enhance our understanding of the SPT phase transitions.

A natural question is how one can sharply characterize SPT phases 
in the absence of any local order parameter.
A standard approach is to make use of \textit{non-local} operators \cite{Wen:1990aa,Pollmann:2012ab,Haegeman:2012aa}.
If such an operator takes a nontrivial value in the ground state,
it implies that the system belongs to a nontrivial SPT phase.
Indeed, a class of string order parameters has been shown to be a good non-local order parameter of one-dimensional SPT phases protected by spin-rotation symmetries \cite{Nijs:1989aa,Lou:2003aa,Duivenvoorden:2012aa,Ueda:2014aa,Ogino:2022aa}.
However,
since their ground-state expectation values are not quantized,
SPT phases are not sharply distinguished when the expectation values are small or when the number of phases is larger than two; in the latter case appropriate order parameters are expected to take \textit{complex} numbers.
A symmetry-protected $\mathbb{Z}_2$ Berry phase defined through an adiabatic transformation of a certain class of valence-bond-solid (VBS) states turns out to be quantized \cite{Motoyama:2018aa,Kariyado:2018aa}, but
calculating the $\mathbb{Z}_2$ phase is difficult for general states other than the VBS states since it is not directly obtained from a single wave function.
Furthermore, while exactly quantized order parameters or $\mathbb{Z}_2$ indices have also been found in infinitely long spin chains \cite{Tasaki:2018aa,Ogata:2018aa,Ogata:2020ab,Ogata:2020aa,Tasaki:2023aa},
it is unclear how to numerically calculate such indices for finite chains with periodic boundary conditions (PBCs) \cite{Nakamura:2002aa}, which are common in condensed matter physics.
In addition, efforts have been made to define quantized order parameters using twisting boundary conditions (TBCs) in topological response theories \cite{Yao:2020PRX,Yao:2021aa},
but this strongly relies on the highly nontrivial assumption that the gap is not closed by TBCs~\cite{Oshikawa:2000aa,Watanabe:2018aa} and is not rigorous.

In this Letter,
we show quantization of the non-local order parameter
 $\langle\exp(2\pi\mathrm{i}\sum_{m=1}^Lm\hat{S}^z_m/L)\rangle$ 
in the thermodynamic limit $L\rightarrow\infty$ for spin-$S$ chains preserving U$(1)_z\rtimes\mathbb{Z}^x_2$-spin rotation symmetry under PBCs.
The averaged operator therein is called Lieb-Schultz-Mattis (LSM) twisting operator and was introduced to understand quantum ingappabilities known as the LSM theorem \cite{Lieb:1961aa,Affleck:1986aa,OYA1997,Oshikawa:2000aa,Hastings:2004ab,NachtergaeleSims}.
Before our proof of quantization of the nonlocal order parameter,
it was used to numerically detect VBS order in various VBS states \cite{Nakamura:2002aa}; its sign gives a diagnosis of two distinct phases \cite{Tasaki:2018aa}, while the sign of complex-valued order parameters is not applicable for distinguishing more than two SPT phases. 

In addtion to rigorously proving the quantization,
we apply it to show that the ``magnetic'' lattice translation symmetry must map one 
U$(1)_z\rtimes\mathbb{Z}^x_2$-SPT phase to another \textit{distinct} SPT phase for half-integral spin chains.
Here ``magnetic'' means that the translation symmetry is accompanied by an anti-unitary time reversal.
We also apply this SPT viewpoint to generalize LSM theorem to magnetic translations.
Moreover, we extend our proof of quantization to SU$(N)$ spin chains~\cite{Greiter:2007aa,Fuhringer:2008aa,Duivenvoorden:2012aa,Dufour:2015aa,Tanimoto:2015aa,Nataf:2016aa,Weichselbaum:2018aa,Capponi:2020aa},
where the corresponding nonlocal order parameter is a complex number, and translation symmetry leads to an $N$-ality transformation among distinct SPT phases.
The nonlocal order parameter is quantized to be an $N$-th root of unity and can be used to identify $\mathbb{Z}_N$ SPT phases.
Furthermore,
our rigorous theorems give an efficient approach to construct possible multi-critical phase transitions starting from a single Hamiltonian with a unique gapped ground state.
For example, our construction can be useful in understanding criticality in cold-atomic experimental realizations of SU$(N)$ spin~\cite{Honerkamp:2004aa,Gorshkov:2010aa,Scazza:2014aa,Zhang:2014aa}, and
provide a perspective from SPT phases on the potential stability of multi-criticality observed in their phase diagram.

\paragraph{Setting and Results.---}
We study a spin-$S$ chain of length $L$ under PBCs, described by Hamiltonian $\mathcal{H}$ possessing $\text{U}(1)_z\rtimes \mathbb{Z}^x_{2}\subset\mathrm{SO}(3)$ symmetry with a unique gapped ground state $|\text{G.S.}\rangle$. 
Here, $\text{U}(1)_z$ is continuous symmetry of spin rotation around the $z$-axis generated by $\hat{\bm{S}}^z = \sum_{j=1}^{L} \hat{S}_j^z$,  and $\mathbb{Z}^x_2$ is discrete global symmetry of $\pi$-angle rotation $R_{\pi}^{x}$ around the $x$-axis.
{Here $\hat{S}_j^z$ is a spin operator on site $j$ in an irreducible representation of SU(2).
A more general situation will be considered later for SU($N$) spin models with $N\geq2$ below Eq.~(\ref{SUN_relationship}).}
These two symmetry generators satisfy the relation 
\begin{eqnarray}\label{SU2_relationship}
{\left(R_\pi^x\right)}^{-1}\hat{S}_j^zR_\pi^x+\hat{S}_j^z=0.
\end{eqnarray}

We impose the locality of the Hamiltonian so that
we can decompose $\mathcal{H}$ as a sum of local terms: 
\begin{eqnarray}\label{locality}
\mathcal{H}=\sum_{j=1}^{L}q_j,
\end{eqnarray}
where $q_j$ is localized around site $j$, $i.e.$, it has a nontrivial action at sites within a finite distance $\ell$ from site $j$ and acts as an identity operator at the other sites. 
The maximal interaction range $\ell$ is assumed to be $j$-independent.
Systematic symmetry averaging of $q_j$ yields
\begin{equation}
h_j=\sum_{k=0}^1\int_0^{2\pi}\frac{\mathrm{d}\theta}{4\pi}(R^x_\pi)^k\exp(-\mathrm{i}{\hat{\bm{S}}^z}\theta)q_j\exp(\mathrm{i}{\hat{\bm{S}}^z}\theta)(R^x_\pi)^{-k},
\end{equation}
with which we redefine the Hamiltonian
\begin{equation}
\mathcal{H}=\sum_{j=1}^Lh_j.
\end{equation}
Since $U(1)_z\rtimes\mathbb{Z}^x_2$ is onsite symmetry,
$h_j$'s are localized around site $j$
and respect the \textit{global} symmetry:
\begin{eqnarray}\label{local_sym}
[{\hat{\bm{S}}^z},h_j]=[R^x_\pi,h_j]=0.
\end{eqnarray}
{The Hamiltonian $\mathcal{H}$ implicitly depends on $L$,
and the thermodynamic limit $L\rightarrow\infty$ is taken on such a series of $L$-dependent Hamiltonians,
{with $L=2M$ for half-integer spin chains ($M\in\mathbb{N}$) and no such constraint for integer spin chains} so that the ground state is unique while taking the limit $L\to\infty$, i.e., $M\to\infty$.
We assume that the local terms $h_j$'s are \textit{uniformly} bounded operators ---
for all $h_j$'s, a common (lower and upper) $L$-independent spectrum bound can be found.
This assumption is satisfied by generic spin models.
}

We state our main result as follows: 

\paragraph{Theorem 1:} 
If the Hamiltonian $\mathcal{H}$ has $(\text{U}(1)\rtimes\mathbb{Z}_{2})$-rotation symmetry and a unique gapped ground state $|\text{G.S.}\rangle$ under PBCs, then 
\begin{eqnarray}
\mathcal{I}\equiv\lim_{L\rightarrow\infty}\langle\text{G.S.}|\hat{U}|\text{G.S.}\rangle=\pm1,
\end{eqnarray}
where the operator $\hat{U}$ is exactly the LSM twisting operator \cite{Lieb:1961aa,Nakamura:2002aa}
\begin{eqnarray}\label{TwistingOperator}
\hat{U} = \exp \!\left( \frac{2 \pi \mathrm{i}}{L} \sum_{m = 1}^{L} m \hat{S}^z_m \right).
\end{eqnarray}

\paragraph{Proof of Theorem 1.---} 
Since $|\text{G.S.}\rangle$ is a unique ground state of $\mathcal{H}$, it must also be an eigenstate of {$\hat{\bm{S}}^z$} and $R^x_\pi$,
\begin{equation}
{\hat{\bm{S}}^z}|\text{G.S.}\rangle=0, \qquad
R^x_\pi|\text{G.S.}\rangle=e^{\mathrm{i}\alpha}|\text{G.S.}\rangle,
\label{alpha}
\end{equation}
where $\alpha\in\mathbb{R}$.

Under PBCs we can introduce an equivalence relation for spins {$\hat{S}^z_{m+L}\equiv\hat{S}^z_m$}, which allows us to extend the site index $m$ to arbitrary integers. 
Hence, without loss of generality, we can define more general twisting operators for $j \in [-L, L)$ as follows:
\begin{eqnarray}\label{TwistingOperator_general}
\hat{U}^{(j)}&\equiv&\exp \!\left( \frac{2 \pi \mathrm{i}}{L} \sum_{m = j+1}^{j+L} m \hat{S}^z_m \right)\nonumber\\
&=&\left\{\begin{array}{ll}\hat{U}\exp\left(\sum_{n=1}^j2\pi \mathrm{i}\hat{S}^z_n\right),&\text{ if }j> 0;\\
\hat{U},&\text{ if }j= 0;\\
\hat{U}\exp\left(-\sum_{n=j+1}^{0}2\pi\mathrm{i}\hat{S}^z_n\right),&\text{ if }j< 0,\end{array}\right.
\nonumber\\
&\propto&\hat{U},
\end{eqnarray}
up to a $j$-dependent U(1) phase in comparison with the original definition in Eq.~\eqref{TwistingOperator},
since $2\pi$-rotation of any single spin {in an irreducible representation is physically an identity operator.}

First, we prove $|\mathcal{I}| = 1$.
Let us consider $|\Phi\rangle\equiv\hat{U}|\text{G.S.}\rangle$. 
Similar to the original proof of the LSM theorem \cite{Lieb:1961aa},
we focus on the difference $\Delta E=\langle\Phi|\mathcal{H}|\Phi\rangle-\langle\text{G.S.}|\mathcal{H}|\text{G.S.}\rangle$ of the energy expectation value between $|\text{G.S.}\rangle$ and $|\Phi\rangle$.
The proportionality Eq.~\eqref{TwistingOperator_general} implies that
\begin{eqnarray}
\langle\Phi|\mathcal{H}|\Phi\rangle=\sum_{j=1}^{L}
 \big\langle \mathrm{G.S.} \big| \hat{U}^{(j-2\ell) \dagger} h_j \hat{U}^{(j-2\ell)} \big| \mathrm{G.S.} \big\rangle,
\end{eqnarray}
which avoids the coordinate jump in $\hat{U}$ near $j$ due to the PBCs. 
We further make use of the global U$(1)_z$ transformation $\exp(-2\pi\mathrm{i}\hat{S}^zj/L)$ \textit{within each $j$-summand} and the locality of $h_j$'s to obtain
\begin{eqnarray}
\Delta E &=& \sum_{j=1}^{L}\biggl\langle\mathrm{G.S.}\bigg|\exp\!\left[-\frac{2\pi\mathrm{i}}{L}\!\sum_{|m-j|\leq \ell}(m-j)\hat{S}^z_m\right]\! h_j\nonumber\\
&& \times\exp \!\left[ \frac{2\pi\mathrm{i}}{L}\!\sum_{|n-j|\leq \ell}(n-j)\hat{S}^z_n\right]\! -h_j\bigg|\mathrm{G.S.}\biggr\rangle . \quad 
\end{eqnarray}
{Since $h_j$'s are uniformly bounded and the coefficients $m-j, n-j$ of $\hat{S}^z_m$ and $\hat{S}^z_n$ are restricted in $[-\ell, \ell]$, thereby bounded independently of $L$,} 
the exponentials can be expanded as a Taylor series of $(1/L)$:
\begin{eqnarray}
\Delta E&=&\sum_{j=1}^{L}\frac{2 \pi \mathrm{i}}{L}  \biggl\langle \mathrm{G.S.} \bigg| \left[h_j, \sum_{|m-j|\leq \ell} (m-j) \hat{S}^z_{m} \right] \bigg| \mathrm{G.S.} \biggr\rangle \nonumber\\
&&+\sum_{j=1}^{L} \langle \mathrm{G.S.} | \mathscr{O} (1 / L^2) | \mathrm{G.S.} \rangle. 
\end{eqnarray}
Then we use the $R_\pi^x$ symmetries of $h_j$ and $|\text{G.S.}\rangle$ to replace $\hat{S}^z_{m}$ by $[\hat{S}^z_{m} +(R_\pi^x)^\dagger \hat{S}^z_{m}R_\pi^x]/2$ on the first line,
which vanishes using Eq.~\eqref{SU2_relationship}. 
Consequently, the leading contributions come from the second line, 
$\Delta E =\mathscr{O} (1 / L)\rightarrow0$.

Given the condition of \textit{Theorem 1} that $|\text{G.S.}\rangle$ is the unique, gapped ground state of $\mathcal{H}$,
we conclude from vanishing $\Delta E\to0$ that
$|\mathcal{I}|=\lim_{L\rightarrow\infty}|\langle\text{G.S.}|\Phi\rangle|=1$,
i.e., $\mathcal{I}\in\text{U}(1)$, or equivalently, we have
\begin{eqnarray}
\label{SU2_module1}
\lim_{L\rightarrow\infty}|\Phi\rangle=\lim_{L\rightarrow\infty}\hat{U}|\text{G.S.}\rangle=\mathcal{I}|\text{G.S.}\rangle.
\end{eqnarray}
This means that
$|\text{G.S.}\rangle$ is an asymptotic eigenstate of $\hat{U}$ in the thermodynamic limit.

Let us suppress various limit symbols in the following discussion, since the limit of the product of interest can be taken in each multiplicand.
From the identity that follows from Eq.~(\ref{SU2_relationship}),
\begin{eqnarray}\label{identity}
1=\bigl[(R^x_\pi)^\dagger \hat{U}R^x_\pi\bigr]\hat{U},
\end{eqnarray}
we obtain
\begin{eqnarray}\label{SU2_square1}
1&=&
\langle\text{G.S.}|\bigl[(R^x_\pi)^\dagger \hat{U}R^x_\pi\bigr]\hat{U}|\text{G.S.}\rangle
\nonumber\\
&=&
\langle\text{G.S.}|e^{-\mathrm{i}\alpha}\hat{U}R^x_\pi \mathcal{I}|\text{G.S.}\rangle
\nonumber\\
&=&
\mathcal{I}\langle\text{G.S.}|e^{-\mathrm{i}\alpha}\hat{U}e^{\mathrm{i}\alpha}|\text{G.S.}\rangle=\mathcal{I}^2,
\end{eqnarray}
where we have used the property that $|\text{G.S.}\rangle$ and $|\Phi\rangle$ are both asymptotically the unique ground state of $\mathcal{H}$, thereby an asymptotic eigenstate of $R^x_\pi$ symmetry {with the eigenvalue $e^{\mathrm{i}\alpha}$ as in Eq.~\eqref{alpha}}.
Hence $\mathcal{I} \in \{\pm 1\}$, and \textit{Theorem 1} follows.

\paragraph{SU(N) case:---} 
We notice that the essential ingredients for deriving \textit{Theorem 1} are Eqs.~(\ref{SU2_relationship}) and  (\ref{local_sym}), which come from SU$(2)$ algebra and the locality.
Thus,
the quantization is readily generalized to SU$(N)$ spins if the counterparts of $\mathcal{I}$ and the two equations are provided.
The generators of $\mathbb{Z}_N$ and U$(1)$ in the fundamental representation of SU$(N)$, corresponding to $R^x_\pi$ and $\hat{S}_j^z$ of SU(2) respectively, are given by
\begin{equation}\label{SUN_generator}
	R_{N}^{x} = 
	\begin{pmatrix}
		\ & 1 \\
		\mathbb{I}_{N-1} & \ 
	\end{pmatrix}
	, \  
	\hat{t}^z_{{N,j}} = \frac{1}{N}
	\begin{pmatrix}
		\mathbb{I}_{N-1} & \ \\
		\ & -(N-1) 
	\end{pmatrix},
	\end{equation}
where $\mathbb{I}_{N-1}$ is $(N-1)$-dimensional identity matrix. 
Similarly, the relationship Eq.~\eqref{SU2_relationship} is reformulated as
\begin{eqnarray}\label{SUN_relationship}
\hat{t}^z_{{N,j}}+\sum_{k=1}^{N-1} ({R_{N}^{x \ \dagger}})^{k}\hat{t}^z_{{N,j}}(R_{N}^{x})^{k}=0,
\end{eqnarray}
which holds for other representations since any irreducible representation must be a sub-representation of a tensor product of several fundamental representations.
{In fact, we can consider the case where $\hat{t}_{N,j}^z$'s are in any irreducible representation,
or, more generally, any direct sum of irreducible representations that have the same number $b$ of Young-tableau boxes modulo $N$.
This ensures the generalization of the proportionality in Eq.~(\ref{TwistingOperator_general}).
Additionally, the thermodynamic limit will be taken as $L=MN/\text{gcd}(b,N)$ with $M\rightarrow+\infty$, where ``gcd'' denotes the greatest common divisor.}

The symmetry of interest is
the smallest subgroup of PSU$(N)=\text{SU}(N)/\mathbb{Z}_N$ ``spin''-rotation group that contains $\text{U}(1)\cup \mathbb{Z}_{N}$.
{This symmetry is $[\text{U}(1)]^{N-1}\rtimes\mathbb{Z}_N\subset\text{PSU}(N)$ and 
the symmetry averaging procedure to arrive at Eq.~(\ref{local_sym}) for $N>2$ can be done by using its semi-product structure,
as explained in~\cite{SupMat}.}
Then the following theorem for SU$(N)$ spin chains can be derived by repeating the derivation of \textit{Theorem 1}.

\paragraph{Theorem 2:} 
If {$G\equiv[\text{U}(1)]^{N-1}\rtimes\mathbb{Z}_N$} is respected by SU($N$)-spin Hamiltonian with a unique gapped ground state,
then:  
\begin{eqnarray}
\mathcal{I}_N\in\{\omega^k|k=0,1,\cdots,N-1\},
\end{eqnarray}
where $\omega\equiv\exp(2\pi\mathrm{i}/N)$ and
\begin{eqnarray}
\mathcal{I}_N\equiv\lim_{L\rightarrow\infty}\langle\text{G.S.}|\hat{U}_N|\text{G.S.}\rangle, 
\end{eqnarray}
with the generalized twisting operator
\begin{eqnarray}
\hat{U} _N= \exp \!\left( \frac{2 \pi \mathrm{i}}{L} \sum_{m = 1}^{L} m \hat{t}_{N,m}^z \right).
\end{eqnarray}
Here the quantization is obtained through the identity
\begin{equation}
1=\bigl\{[(R^x_N)^\dagger]^{N-1} \hat{U}_N(R^x_N)^{N-1}\bigr\}\cdots\bigl[(R^x_N)^\dagger \hat{U}_NR^x_N\bigr]\hat{U}_N,
\end{equation}
which follows from Eq.~(\ref{SUN_relationship})
and extends Eq.~(\ref{identity}).

\paragraph{Applications to SPT phases.---}
Although the quantization by \textit{Theorems 1} \& \textit{2} implies that $\mathcal{I}_N$ can distinguish two different $G$-SPT phases,
for this order parameter to be useful, it must be proven that at least two possible $\mathcal{I}_N$ values can be realized.
In fact, $\mathcal{I}_N$ can exhaust all the possible values:\par
\paragraph{Corollary 3:} 
SU($N$) spin chains have at least $N$ distinct $[\text{U}(1)]^{N-1}\rtimes\mathbb{Z}_N$-symmetric SPT phases that are characterized by all different values of $\mathcal{I}_N$ and related to each other by lattice translation,
provided that $N$ is coprime to the Young-tableau box number of SU$(N)$ spins.

Let us illustrate this for the $N=2$ case where Hamiltonian $\mathcal{H}$ of a spin-$\frac12$ chain has a unique, gapped ground state $|\text{G.S.}\rangle$ and the order parameter $\mathcal{I}$. 
We consider another gapped Hamiltonian $\mathcal{H}'\equiv T\mathcal{H}T^{-1}$ with $\mathcal{I}'$,
where $T$ is one-site translation operator acting on spins as $T^{-1}\vec{S}_jT=\vec{S}_{j+1}$ (under PBCs).
$\mathcal{H}'$ has the ground state $|\text{G.S.}'\rangle=T|\text{G.S.}\rangle$. 
By $\mathcal{I}'=\langle\text{G.S.}'|\hat{U}|\text{G.S.}'\rangle$,
\begin{eqnarray}\label{permute}
\mathcal{I}'&=&\left\langle\text{G.S.}\bigg|\hat{U} \exp (2 \pi \mathrm{i} \hat{S}^z_1) \exp \left(- \frac{2 \pi \mathrm{i}}{L} \hat{\bm{S}}^z \right)\bigg|\text{G.S.}\right\rangle\nonumber\\
&=&(-1)^{2S}\mathcal{I} = -\mathcal{I},
\end{eqnarray}
where $\hat{\bm S}^z|\text{G.S.}\rangle=0$ due to the $\mathbb{Z}_2$ symmetry and $\exp(2\pi \mathrm{i}\hat{S}_1^z)=(-1)^{2S}=-1$ for $S=\frac12$. 
It also means that any half-integral-spin $G$-SPT Hamiltonian and its image Hamiltonian of translation $T$ must belong to two distinct SPT phases. 
Moreover,
the translation transformation $T$ above can be replaced with
\begin{equation}\label{antiunitary}
\widetilde{T}=T\circ\hat{Q}\text{ or } \widetilde{T}=\hat{\Theta}\circ T\circ\hat{Q},
\end{equation}
where $\hat{\Theta}$ is the \textit{antiunitary} spin time-reversal transformation, and $\hat{Q}$ denotes any other unitary transformation which commutes with $\hat{S}_j^z$.
Such a combination of translation with anitunitary symmetry is called magnetic translation. 

The above proof can be extended to $N>2$ cases with $\exp(2\pi\mathrm{i}\hat{S}^z_1)=(-1)^{2S}$ replaced by
\begin{equation}
\exp(2\pi\mathrm{i} \hat{t}_{N,1}^z)=\omega^b,
\end{equation}
leading to
\begin{equation}\label{shift}
\mathcal{I}_N'={\omega^b}\,\,\mathcal{I}_N,
\end{equation}
where $b$ is the Young-tableau box number of the SU$(N)$ ``spin'' \cite{Georgi:2000} on the first site. Since $b=1$ for the fundamental representation,
we conclude that $T$ is an $N$-ality transformation connecting $N$ distinct $G$-SPT phases.
Unfortunately,
when $N>2$,
there is no analog of antiunitary time-reversal that anticommutes with $\hat{t}^z_N$,
so only $\widetilde{T}=T\circ\hat{Q}$ can be realized.

\paragraph{Applications to LSM ingappabilities and multi-phase transitions.---}
It is worth highlighting that our theorem can also be an important tool for obtaining the quantum ingappability, as shown by the following corollary:
\paragraph{Corollary 4:}
Any $[\text{U}(1)]^{N-1}\rtimes\mathbb{Z}_N$-symmetric and $\widetilde{T}$-symmetric Hamiltonian of SU$(N)$ spin with $b$ boxes ($b$-spin for short) per unit cell cannot have a unique, gapped ground state, if $b$ is an integer coprime to $N$.

\textit{Corollary 4} is proven by contradiction. 
Suppose that a Hamiltonian $\mathcal{H}$, which is $G$-symmetric and $\widetilde{T}$-symmetric with a $b$-spin per unit cell, has a unique, gapped ground state.
Then $\mathcal{I}_N'=\mathcal{I}_N$ since $\mathcal{H}'\equiv\widetilde{T}\mathcal{H}\widetilde{T}^{-1}=\mathcal{H}$,
which contradicts with Eq.~\eqref{shift}.
\textit{Corollary 4} is exactly an extension of the LSM theorem in terms not only of the type of ``spins'', but also of the translation symmetry.
The original proof of LSM theorem strongly relies on the \textit{unitarity} of the translation to have a well-defined lattice momentum,
which is unavailable for antiunitary $\widetilde{T}$.

Furthermore,
\textit{Theorem 2} gives an approach to construct a potential multi-critical phase transition when $b\nmid N$.
We define $k\equiv N/\text{g.c.d.}(b,N)$, where ``g.c.d.'' denotes the greatest common divisor.
We infer from the second half of \textit{Corollary 3} and its proof that Hamiltonian defined by
\begin{equation}
\mathcal{H}(s_1,\cdots,s_k)\equiv \sum_{j=1}^{k}s_j\widetilde{T}^{j-1}\mathcal{H}\widetilde{T}^{-(j-1)},\,\,s_j\in[0,1],
\end{equation}
can be a potential candidate to describe a $m$-critical phase transition with $m\leq k$.
When $N=2$,
$k$ can be only $1$ or $2$.
We can take $\widetilde{T}=T$ and $\mathcal{H}=\sum_j[1+(-1)^j]\vec{S}_j\cdot\vec{S}_{j+1}$, and the critical {line $s_1=s_2$} is in the SU$(2)_1$ universality class~\cite{Affleck:1987aa,Affleck:1988aa}.
The phase diagram will be much richer when $N>2$,
and $N$-ality of $\widetilde{T}$ provides a systematic way to discover multi-critical points starting from a \textit{single} SPT Hamiltonian $\mathcal{H}$.
In addition,
for a generic $\mathcal{H}$,
$\mathcal{H}(s_1,\cdots,s_k)$ can break $\widetilde{T}$ symmetry for any $\{s_j\}$,
so the inevitable criticalities are beyond the LSM ingappabilities of \textit{Corollary 4}.
This implies the potential stability of critical phases, which can be observed numerically or experimentally~\cite{Greiter:2007aa,Fuhringer:2008aa,Duivenvoorden:2012aa,Dufour:2015aa,Tanimoto:2015aa,Nataf:2016aa,Weichselbaum:2018aa,Capponi:2020aa}.

\paragraph{Conclusions and discussions.---}
In this work, we have proposed a precisely quantized topological order parameter for SPT phases with $[\text{U}(1)]^{N-1}\rtimes\mathbb{Z}_N$ symmetry. 
The lattice translation induces the $N$-ality transformation connecting $N$ distinct SPT phases for SU$(N)$ spin chains in the fundamental representation.
Moreover, the translation transformation is also extended through the incorporation of \textit{antiunitary} transformation in the SU(2) case where $S$ is a half integer, and 
the extended LSM ingappabilities by magnetic translations follow. 
The quantization of the topological order parameter is expected to be useful for diagnosing phase transitions between, e.g., a valence-bond phase \cite{Rao:2014aa} and a chiral spin liquid \cite{Sorella:2003aa}, those at one-dimensional deconfined quantum critical points \cite{Mudry:2019,Zheng:2022aa},  or those 
in complicated SU$(N)$ spin models \cite{Duivenvoorden:2012aa,Morimoto:2014aa,Weichselbaum:2018aa,Gozel:2020aa,Capponi:2020aa}
and even in some experimental realizations \cite{Li:2021aa}.
In addition,
the $N$-ality property of translation symmetry provides an efficient way to construct possible multi-critical phase transitions from a single uniquely gapped spin Hamiltonian.
The study of such multi-criticalities, universalities and exotic emergent symmetries can be of future interest.

\begin{acknowledgements}
\paragraph{Acknowledgements.---}
The authors thank Linhao Li, Masaki Oshikawa and Yasuhiro Tada for helpful discussions. 
Y.~Y. thanks the sponsorship from Yangyang Development Fund and Xiaomi Young Scholars Program.
The work of A.F.\ was supported in part by JSPS KAKENHI (Grant No.~JP19K03680) and JST CREST (Grant No.~JPMJCR19T2).
\end{acknowledgements}

\bibliographystyle{apsrev4-1}

%

\onecolumngrid 
\appendix
\section{Supplemental Materials: Symmetrization of local Hamiltonian terms}

In this part, we generally deal with $N \geq 2$ cases, and the Hamilitonian $\mathcal{H}$ respects the symmetry $G$ generated by U$(1)$ and $\mathbb{Z}_N$. 
Locality means that $\mathcal{H}$ can be decomposed as
\begin{eqnarray}
\mathcal{H}=\sum_jq_j,
\end{eqnarray}
where $q_j$ can act nontrivially only on the sites within coordinates $[j-l,j+l]$ and $l$ is $j$-independent.

The symmetry group $G$ generated by U(1) and $\mathbb{Z}_N$ is actually a semi-direct product of $\mathbb{U}\rtimes\mathbb{Z}_N$, where
\begin{eqnarray}\label{mU}
\mathbb{U}&=&\left\{u_{\vec{\alpha}}\equiv\exp\left[\sum_{n=0}^{N-2}\mathrm{i}\alpha_n{\left[\left(R_{N}^{x}\right)^{-n}\hat{t}_N^z\left(R_{N}^{x}\right)^n\right]}\right]:\alpha_n\in[0,2\pi N)\right\},\\
&\cong&[\text{U}(1)]^{N-1},
\end{eqnarray}
where $\vec{\alpha}\equiv[\alpha_0,\alpha_1,\cdots,\alpha_{N-2}]$.
One can also prove that
\begin{eqnarray}
G\cong\bigsqcup_{k=0}^{N-1}\mathbb{U}\left(R_{N}^{x}\right)^{k},
\end{eqnarray}
by which
we can symmetrize $q_j$ to be
\begin{eqnarray}
h_j=\sum_{k=0}^{N-1}\frac{1}{N}\prod_{n=0}^{N-2}\int_0^{2\pi N}\frac{\mathrm{d}\vec{\alpha}}{2\pi N}{\left[u_{\vec{\alpha}}\left(R_{N}^{x}\right)^k\right]}q_j{\left[u_{\vec{\alpha}}\left(R_{N}^{x}\right)^k\right]^\dagger},
\end{eqnarray}
which is a new and \textit{globally} symmetric decomposition of $\mathcal{H}$:
\begin{eqnarray}
&&\mathcal{H}=\sum_jh_j,\\
&&[h_j,\hat{t}^z_N]=[h_j,R^x_N]=0,
\label{Symmetry_of_hj}
\end{eqnarray}
{which means that the local decomposition $h_j$ respects the global $G$ symmetry}.
Furthermore, $h_j$ is also a local term with interaction range $2l$ around the site $j$ since $[u_{\vec{\alpha}}${$\left(R_{N}^{x}\right)^k]$} is onsite thereby locality preserving.

\end{document}